\documentclass[manuscript]{acmart}

\usepackage{xspace}
\usepackage{hyperref} 
\usepackage{dsfont}

\newcommand{\shortorextended}[2]{{\color{black} #2}} 

\newcommand{\extended}[1]{{\shortorextended{}{#1}}} 

\usepackage{todonotes}
\presetkeys
    {todonotes}
    {backgroundcolor=green}{}

\newcommand{\ie}{i.\,e.,\xspace}
\newcommand{\eg}{e.\,g.,\xspace}

\AtBeginDocument{
  \providecommand\BibTeX{{
    \normalfont B\kern-0.5em{\scshape i\kern-0.25em b}\kern-0.8em\TeX}}}

\copyrightyear{2021}   
\acmYear{2021}   
\setcopyright{acmlicensed}\acmConference[iiWAS2021]{The 23rd International Conference on Information Integration and Web Intelligence}{November 29-December 1, 2021}{Linz, Austria}
\acmBooktitle{The 23rd International Conference on Information Integration
and Web Intelligence (iiWAS2021), November 29-December 1, 2021, Linz, Austria}
\acmPrice{15.00}
\acmDOI{10.1145/3487664.3487712}
\acmISBN{978-1-4503-9556-4/21/11}

\usepackage{graphicx}
\usepackage{textcomp}
\usepackage{xcolor}
\usepackage{algorithm}
\usepackage{xcolor}
\usepackage{colortbl}

\usepackage{algpseudocode}

\def\BibTeX{{\rm B\kern-.05em{\sc i\kern-.025em b}\kern-.08em
    T\kern-.1667em\lower.7ex\hbox{E}\kern-.125emX}}

\newcommand{\degreeOfFreedom}{\text{degreeOfFreedom}}
\newcommand{\statisticVal}{\text{statisticVal}}
\newcommand{\pvalue}{\text{pvalue}}
\newcommand{\topic}{\text{topic}}
\newcommand{\conditions}{\text{conditions}}


\begin{document}

\title[
\shortorextended{STEREO:}{} A Pipeline for Extracting Experiment Statistics, Conditions, and Topics
]{STEREO: A Pipeline for Extracting Experiment Statistics, Conditions, and Topics from Scientific Papers}

\author{Steffen Epp}
\author{Marcel Hoffmann}
\author{Nicolas Lell}
\author{Michael Mohr}
\author{Ansgar Scherp}
\affiliation{\institution{\\Ulm University}\country{Germany}
}
\email{{steffen.epp, marcel.hoffmann, nicolas.lell, michael.mohr, ansgar.scherp}@uni-ulm.de}

\begin{abstract}
\extended{We address the problem of extracting reports of statistics along with information about the experiment conditions and experiment topics from scientific publications.}
A common writing style for statistical results are the recommendations of the American Psychology Association (APA).
In practice, writing styles vary as reports are not 100\% following APA-style or parameters are not reported despite being mandatory. In addition, the statistics are not reported in isolation but in context of experiment conditions investigated and the general experiment topic. We address these challenges by proposing a flexible pipeline STEREO based on wrapper induction and unsupervised aspect detection to extract experiment statistics, conditions, and topics.
Thus, in contrast to existing rule-based tools like statcheck with a pre-defined set of rules, we learn rules via induction.
\extended{Hierarchical wrapper induction is applied to learn rules to extract the reported statistics.
Challenge here is to apply wrapper induction on an information extraction task without having formatting landmarks as they can be exploited in HTML pages.
Result of step 1 is a set of extracted statistic reports together with sentences in which the reports were found.
This is used as input to step 2 of STEREO, which has two two parts.
We extract experiment conditions using a grammar-based wrapper.
Furthermore, we identify the experiment topic using an unsupervised attention-based aspect extraction approach adapted to our problem domain.
We applied our pipeline to the over $100,000$ documents in the CORD-19 dataset.}
It required only $0.25\%$ of the CORD-19 corpus (about $500$ documents) to learn statistics extraction rules that cover $95\%$ of the sentences in CORD-19. The statistic extraction has $100\%$ precision on APA-conform statistics, which is identical with statcheck. In addition, STEREO can extract non-APA writing styles with $95 \%$ precision, which statcheck does not support. Extracting non-APA conform statistics is important as they make more than $99\%$ of all $113$k extracted statistics. We could extract in $46\%$ the correct conditions from APA-conform reports ($30\%$ for non-APA). The best model for topic extraction achieves a precision of $75\%$ on statistics reported in APA style ($73\%$ for non-APA conform).
\extended{We conclude that STEREO is a good foundation for automatic statistic extraction and future developments for scientific paper analysis.
Particularly the extraction of non-APA conform reports is important and allows applications such as giving feedback to authors about what is missing and could be changed.
Finally, STEREO complements the portfolio of metadata extraction tools and can be integrated in a general scientific paper analysis pipeline.

\vspace{5mm}

\textbf{Note: This is an extended version of our earlier iiWAS2021 paper. Please cite the paper given below.}

}
\end{abstract}


\maketitle

\section{Introduction}
\label{ch:Introduction}

In many fields of science the research results are analyzed and presented with statistical methods, e.\,g., in psychology, life sciences, social sciences, economics, 
and others.
Therefore, there is a large amount of scientific papers which contain statistical data in an unstructured way. 
In addition, statistics are not reported in isolation but together with the experiment conditions and experiment topics.
Our objective is to identify and extract such data from scientific papers.
Tools like statcheck~\cite{Statcheck} and its extension~\cite{DBLP:conf/www/LankaR0G21} use a fixed set of regular expressions (rules) to extract statistical reports conform to the American Psychology Association (APA, \url{https://apastyle.apa.org/}).
In contrast, our aim is to extract also reports which are not conform to APA. 
This is needed as in practice writing styles vary, \eg reports are not following APA-style and generally mistakes are being made in reporting statistics.
In fact, our experiments show that of the extracted $113$k statistics, more than $99\%$ of all reported statistics are not conform to APA.
Furthermore, we also extract experiment conditions and experiment topics from scientific papers, which is yet not addressed by the existing tools.

Our pipeline STEREO (STat ExtRaction Experimental cOnditions) uses wrapper induction to find regular expressions for statistics extraction, even if the reporting does not strictly follow APA guidelines.
For example, with these rules we can extract statistics like ``\textit{Physical demand $(t(23) = -2.22, p=0.37)$ and temporal demand $(t(23) = 2.72, p = .012)$ are significantly different}'', although APA style dictates the statistics to be at the end of the sentence and $p$-values are not to be reported with a leading $0$.
Beside the robust extraction of not completely APA-conform statistics, we extract experiment conditions and experiment topics such as ``men'', ``women'' and ``personal\_data'' as reported in the following example: 
``\textit{There was no significant effect for sex, (t(38) = 1.7, p = .097) despite women attaining higher scores than men}''.
This helps to increase the interpretability of our extracted statistical records. 
For extracting the conditions, we apply aspect extraction techniques, namely Attention-based Aspect Extraction (ABAE) \cite{abae}, and grammar-based condition extraction (GBCE). 
The grammar-based approach applies rules based on English grammar and frequently occurring tokens to extract experiment conditions of the corresponding statistic.
Overall, the extracted details of the statistical report, experiment condition, and experiment topics results in a structured metadata record.
For the above example, we extract  $\{ \degreeOfFreedom = 38$, $\statisticVal = 1.7$, $\pvalue = .097$, $\conditions = \{men, women\},~$\topic = personal\_data$ \}$.

We apply and evaluate the STEREO pipeline on the CORD-19 dataset.
For learning the statistic extraction rules, we used $500$ documents, \ie $0.25\%$ of the corpus.
\extended{Currently, the models support the statistics Pearson’s Correlation, Spearman Correlation, Student's $t$-test, ANOVA, Mann-Whitney U Test, Wilcoxon Signed-Rank Test, and Chi-Square Test. 
As it is a wrapper induction approach, it is easy to extend the rule set to other types of statistic and learn corresponding rules.}
Our results show a precision of $100\%$ for the statistic extraction in the case of APA-conform reports, which is equivalent to statcheck's APA style extraction rules~\cite{Statcheck} and its extension, which reported $99\%$ precision on extracting the $p$-value with test statistic~\cite{DBLP:conf/www/LankaR0G21}.
For non-APA conform statistics the precision is $95\%$.
\extended{Furthermore, some statistic types were observed more often than others.
For example, more Pearson correlations than chi-square tests were found.
In addition, it was analyzed how different pairs of parameters were missing.}
STEREO's ability to extract non-APA conform reports is important as it allows to use it in applications like feedback to authors about what is missing and could be changed.

For the extraction of the experiment conditions and experiment topics, the results are mixed as the problem is much more difficult.
Nevertheless, the extraction of experiment conditions has a precision of $46\%$ for APA-conform reports and $30\%$ for non-APA samples.
For topic extraction, we achieve a best precision of $75\%$ on statistics reported in APA style and $73\%$ for non-APA conform statistics.
About half of them are the trivial topic ``statistic'', which is expected given the input data are sentences from the statistics extraction step, but can be easily filtered out from the results.

STEREO can be easily adapted to other datasets and domains.
If writing styles in that domain differ from life sciences, one would use the wrapper induction to add more rules.
The current rule base is already applicable to the range of domains covered by the CORD-19 dataset, as the rules cover $95\%$ of the sentences in the dataset.
Overall, STEREO is a good foundation for automatic extraction of statistics and future developments for scientific paper analysis such as for condition and topic extraction.
STEREO complements the portfolio of existing metadata extraction tools and can be integrated in a general scientific paper analysis pipeline.
An extended version of this paper can be found on arXiv~\cite{DBLP:journals/corr/abs-2103-14124}.

\extended{The paper is organized as follows:}
Below, we discuss works related to our approach. 
In Section~\ref{sec:pip}, we describe the steps of the extraction pipeline and its three components for extracting statistics, experiment conditions, and experiment topics.
The experimental apparatus is described in Section~\ref{sec:expApp} and the results are presented in Section~\ref{sec:results}.
We discuss the results in Section~\ref{sec:disc}, before we conclude.

\section{Related Work}
\label{sec:relatedWork}

First, we discuss general and bibliographic metadata extraction from scientific papers.
Subsequently, we present works for extracting scientific metadata, followed by a presentation of the state of the art in aspect extraction.

\paragraph{General Bibliographic Metadata Extraction}
The extraction of general bibliographic metadata from scientific papers, such as titles, sections or bibliography, is a well studied problem where different solutions are available such as CERMINE~\cite{CERMINE} and Grobid~\cite{GrobidPlain}.
CERMINE is a comprehensive tool for automatic metadata extraction such as title, author, abstracts, and many more. 
Furthermore, it provides the bibliographic references along with their metadata and the full text of the paper, structured in sections and subsections. 
CERMINE has two phases. 
In phase one, it segments the page in meta structures like tiles, sections, and bibliography. 
\extended{To achieve this, all the characters along with their dimensions and coordinates are extracted.} 
Subsequently, the hierarchical structure in pages, zones, lines, words, and characters are extracted by a bottom-up 
algorithm. 
Finally, the document's zones are classified by a support vector machine and rule-based approach into the categories metadata, body, references, and other.
Similar, Grobid\extended{~(GeneRation Of BIbliographic Data)\footnote{\url{https://github.com/kermitt2/grobid}}}
is a framework based on machine learning for extracting, parsing, and re-structuring raw documents such as PDF into structured XML/TEI encoded documents~\cite{GrobidPlain}. 
In contrast to CERMINE, Grobid's machine learning architecture follows a cascade approach and the models are trained using conditional random field (CRF) models.
Each CRF model is optimized on handling different metadata information. 
\extended{The most comprehensive model processes the header information of a scientific paper and extracts different metadata information such as titles, authors, affiliations, address, abstract, keywords, etc.} 
To extract references, $k$-means clustering is used to divide the reference zones into references strings and extract the reference metadata by using a CRF. 
\extended{In the end, the output is a XML document which represents the hierarchical structure of the document and has tags for the extracted metadata.
After a complete processing of a PDF, Grobid created $55$ labels for relatively fine-grained structures, ranging from traditional publication metadata (title, author first/last/middlenames, affiliation types, detailed address, journal, volume, issue, pages, doi, pmid, etc.) to full text structures (section title, paragraph, reference markers, head-/footnotes, figure headers, etc.).}
Similar to these tools, we structure our approach into phases and using a nesting of rule-based and statistical models. 

\paragraph{Scientific Metadata Extraction}
Beyond general purpose metadata extraction tools, there are more specific extraction tools that relate to our work.
For example, Grobid-quantities~\cite{grobidQ} is an extension of Grobid for extracting and normalizing measurements, \ie numerical data from scientific papers and patents. 
The extraction supports quantities (atomic values, intervals, and lists), units (such as length, weight), and different value representations (numeric, alphabetic, or scientific notation).
These extracted measurements are then normalized toward the International Systems of Units (SI). 
The architecture of Grobid-quantities is separated into the steps tokenization, measurement extraction, and parsing and quantity normalization. 
\extended{Before the tokenization step, the text or PDF is structured using Grobid. 
In the tokenization step, the tokens are created by splitting by punctuation marks and is then re-tokenized to separate adjacent digits and alphanumeric characters.
The tokens from the tokenization step are then passed through a cascade of three CRF models, one for quantities, units, and values, respectively.
A list of units with their characteristics is provided for English, German, and French. 
This so called Unit Lexicon is used for labeling. 
For the normalization, an external Java library called Units of Measurement is used.}
The tool statcheck~\cite{Statcheck} uses a fixed set of regular expressions to extract APA-conform reports for common test statistics used in psychology such as $t$, $F$ and $\chi^2$ statistics.
Only statistics written in APA style notation can be extracted with statcheck, \ie it misses any statistic that is written in a slightly different writing style.
The regular expression for each statistics have been hard-coded into the tool.
Once a statistic is extracted, statcheck recomputes the $p$-values to validate the reported statistic.
\extended{Analyzing the actual data distribution for pre-conditions such as type of data (interval vs. ordinal), skewness, or variance is beyond the scope of statcheck.}
Lanka et al.~\cite{DBLP:conf/www/LankaR0G21} extended statcheck by supporting more statistical tests and extracting the sample sizes and number of hypotheses tested. 
In contrast to statcheck and its extension, we do not assume that the reported statistic is perfectly written in APA style.
It is a well known problem that oftentimes crucial information such as the degree of freedom is missing in a reported statistic~\cite{metaStat} or uses a syntax different from APA.
Using a flexible wrapper induction approach, we can learn rules for any writing styles of reported statistics\extended{~and deviations from APA}.

\extended{In a larger context, our work embeds in initiatives such as the Automated Screening Working Group\footnote{\url{https://scicrunch.org/ASWG}}.
The goal of this initiative is to process manuscripts in the biomedical sciences and to provide customized feedback to improve that manuscript, such as an automated screening of COVID-19 preprints~\cite{aswg-covid19-preprint-screening}.
Five metadata extraction tools have been used that extract different information from the papers.
SciScore\footnote{\url{https://www.sciscore.com/}} is a commercial services to extract information on blinding, randomization, sample-size calculations, sex/gender, ethics and consent statements, resources, and Research Resource Identifiers (RRIDs)\footnote{\url{https://www.rrids.org/}}.
Other tools detect the use of open data sets (ODDPub~\cite{DBLP:journals/datascience/RiedelKB20}), 
explicit mentioning of limitations (Limitation-Recognizer~\cite{DBLP:journals/jamia/KilicogluRMR18}), 
visual depictions of data (Barzooka\footnote{\url{https://github.com/NicoRiedel/barzooka}} 
and JetFighter~\cite{jetfighter}), as well domain-specific metadata of 
a correct identification of nucleotide sequences (Seek and Blastn\footnote{\url{http://scigendetection.imag.fr/TPD52/Va/}}).
Thus, statistic extraction and condition extraction as it is considered here has yet not been done in this initiative but is planned to be contributed in the future.}

\paragraph{Aspect Extraction}
\extended{We discuss the related work on aspect extraction, as it is related to our task of detecting and extracting sentence topics and experiment conditions from text.}
The method by Liu et al.~\cite{automatedRS} is an unsupervised approach for selecting optimal rules for aspect extraction.
The rules exploit grammar dependency relations between opinion words and aspects.
The approach aims to effectively select a set of rules automatically. 
Therefore, a small subset of manually selected rules based on a set of dependency relations is used as input.
For this set of rules, the authors' algorithm automatically finds the best subset of rules for the dataset.
\extended{The rules are divided into three types.
The first type of rules is using opinion words to extract aspects, based on dependency relations between them (R1).
The second type of rules is using aspects to extract other related aspects (R2), 
and the third type is using aspects and opinion words to extract new opinion words.
The rule-set selection algorithm runs in three steps.
First, every proposed rule is applied to the training dataset and outputs the precision and recall values of the rule.
For each ruleset R1-R3, a ranking based on the precision of the rules is then calculated.
In step three, leveraging on step~1 and 2, the rules from the ranked rule set are added one by one in descending order and are evaluated.
This is repeated for every rule in the ranked list.
The algorithm then prunes the lower-ranked rules from the rule set to produce the final set of rules only with the best result on the training dataset.}
However, the initial rules need to be carefully selected and tuned manually. 
\extended{This is not possible for our tasks, since we do not have a labeled dataset to classify the usefulness of created rules.}
Xu et al.~\cite{doubleEmb} proposed a method of combining two different word embeddings with a convolutional neural network (CNN) for aspect extraction. 
\extended{Different embeddings and combinations of embeddings with CNNs and long short-term memory (LSTM) based neural networks were tested.} 
It was found, that a general purpose embedding trained on a huge dataset (in their case glove.840B.300d) combined with a domain specific, smaller embedding that is trained for the specific task coupled with a CNN and a final softmax layer performed best.
\extended{Like with Liu et al.~\cite{automatedRS}, we cannot use this approach as it requires a lot of labeled training data.}
Karamanolakis et al.~\cite{studentTeacher} presented a weakly supervised approach for training neural networks for aspect extraction with only a small set of seed words instead of a large labeled training data.
Seed words are keywords describing an aspect that needs to be available for training.
This method adopts the distillation approach~\cite{coreDistillingPaper}, where a simpler neural network (student) gets trained to imitate the predictions of a complex network (teacher). 
\extended{During training, the parameters of the teacher are ``distilled'' to the parameters of the student. 
In the best case, the student will perform comparably to the teacher for the given task but with less complexity.
The teacher is trained on a labeled dataset.} 
As teacher, Karamanolakis et al. used a bag-of-words classifier on the seed words. 
\extended{Therefore, seed words that are predictive of the $K$ aspects get incorporated into (generalized) linear bag-of-words classifiers.}
The student is an embedding-based neural network. 
\extended{First, a segment is embedded and then classified to the $K$ aspects by using the softmax function.}
As embeddings, an unweighted average of Word2Vec (W2V) embeddings~\cite{word2vec} and contextualized BERT embeddings~\cite{BERT} have been used.
The student was trained to imitate the teacher’s predictions by minimizing the cross entropy between the student’s and the teacher’s predictions.
The drawback of this approach is that good seed words are needed for every aspect.
This is possible for aspect extraction on reviews (restaurants, products, etc.), as there is only a known small number of different aspects~\cite{abae}.
In restaurant reviews, for example, two of the aspects the authors mention are price with the seed words price, value, money, worth, and paid, and the second aspect drinks with the seed words wine, beer, glass, and cocktail.
Modeling such aspects with seed words is not feasible for topic extraction, since there can be many different experiment setups with any topic.

Besides the supervised or weakly supervised approaches, He et al.~\cite{abae} proposed an unsupervised Attention-based Aspect Extraction (ABAE) approach. 
ABAE combines word embeddings with an attention mechanism to create sentence embeddings and tries to extract an aspect embedding with an autoencoder. 
ABAE does not need any labels for training. 
One problem is that one has to specify beforehand the number of aspects $K$ that ABEA should try to find in the data.
Finally, Multi-Seed Aspect Extractor (MATE)~\cite{mate} is an extension to ABEA.
MATE uses embeddings of seed words for every aspect to create a seed matrix. 
\extended{By multiplication with a trained or chosen weight vector, these seed matrices are reduced to a vector each and are concatenated to form the aspect matrix. 
This matrix is then used as aspect matrix in ABAE.
Then they use this aspect extraction model together with a polarity prediction model and a segment selection policy to summarize opinions.
In the standard ABAE, the aspect matrix is initialized with the centroids of a clustering on the embedding and then fine-tuned during training.}
MATE's seed method seems to produce slightly better results than standard ABAE. 
But as seed words are needed for every aspect, like in the approach of Karamanolakis et al.~\cite{studentTeacher}, we cannot adopt this method for extracting topics or conditions from scientific experiments.
Thus, we decided to apply ABAE to our problem.
\extended{A detailed explanation will follow in Section~\ref{sec:ABAE}.}

\extended{
\subsection{Summary}

Since the structure of statistical records can differ greatly, the tokenization of them is more complex than, \eg extracting measurements like Grobid-quantities~\cite{grobidQ}.
Thus, we decided for a different approach and use a flexible wrapper induction approach to learn rules to find and extract statistics.
Thus, we have no restrictions on the format of the statistical report except that it can be recognized by humans as a statistic, which contrasts statcheck's APA-style only approach~\cite{Statcheck}.
This way, we are able to learn pattern that can detect statistics which deviate from the APA style guidelines and tolerate incomplete statistics to some degree.
Similarly, we argue for the use of a grammar-based wrapper induction approach for learning rules to extract experiment conditions.
Finally, experiment topics stated in the sentences are extracted with an adaptation of the unsupervised ABEA approach~\cite{abae}.
Main argument here is that in contrast to other aspect extractors like~\cite{mate,studentTeacher}, ABEA's training procedure is fully unsupervised.
This is needed as one cannot provide seed words for all topics and conditions an experiment is about.}

\section{STEREO Pipeline to Extract Experiment Metadata}
\label{sec:pip}

\begin{figure}[h]
    \centering
    \includegraphics[width=0.55\textwidth]{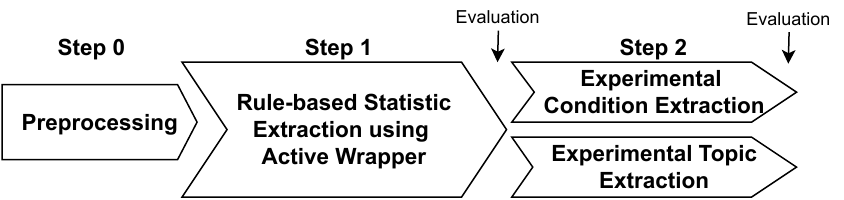}
    \caption{STEREO extraction pipeline to extract the type and values of reported statistics, experiment conditions, and experiment topic. 
    The evaluation points indicate when the steps of the pipeline have been evaluated.}
    \label{fig:pipeline}
\end{figure}

Our metadata extraction pipeline STEREO consists of multiple steps, as illustrated in Figure~\ref{fig:pipeline}.
First, a pre-processing of the input documents is needed, whose challenges and our approach is presented in Section~\ref{sec:preprocessing}.
It takes a set of documents as input and splits them into sentences.
Result of the first step are sentences that can be further processed in the next step to extract statistical information, \ie extract the statistics type and its details as reported in the paper.
Here, an interactive wrapper induction approach is applied which aims to learn rules to extract statistics metadata, which is described in Section~\ref{sec:activeWrap}.
The rules check whether a supported statistics is present in a sentence, and extracts its type and values.
After this step, we have a set of sentences containing reported statistics in plain English as well as corresponding, structured records containing the extracted type and values of the statistic.

This set of sentences and statistics records is given to the final step of our pipeline, which consists of two parallel activities to extract experiment conditions and experiment topics.
Here, two different approaches were taken.
For the extraction of experiment conditions, we base again on a wrapper induction approach. 
However, instead of processing the input sentences as a sequence of characters, our Grammar-based Condition Extraction (GBCE) approach learns its rules on a grammar tree.
The motivation is that the sentences provided by step 1 already contain a report of some statistic and, according to APA style, should also explicitly mention the experiment conditions. 
These mentions of experiment conditions should be identified as noun phrases in the sentences.
The GBCE is described in detail in Section~\ref{sec:gbce}.
For extracting the topic of an experiment, we apply the unsupervised attention-based autoencoder (ABAE) architecture for for aspect extraction~\cite{abae}. 
We adapt ABEA to our purpose of topic extraction as described in Section~\ref{sec:ABAE}.
We provide ABEA a sentence at a time as input, which is then  categorized into a fixed number of aspects.
\extended{The extracted aspects are interpreted as topic.}

\subsection{Preprocessing}
\label{sec:preprocessing}

\extended{Our approach expects a set of documents as separative files in JSON format as used in the CORD-19 dataset, but it can be adapted to any reasonable format.
Tools like Grobid can be used to obtain the correct format, if the files from the given dataset are given in PDF.}
In a first step, the documents are split into sentences.
\extended{The language of each sentence is checked by langdetect\footnote{\url{https://pypi.org/project/langdetect/}}. 
Although our STEREO pipeline is in principal independent of the language, sentences that differs from English will be skipped.
The reason is that for different languages, different rule sets and models have to be learned for the statistics extraction module (step 1) and GBCE and ABEA (step 2).}
\extended{$4.2\%$ of the sentences were removed by the language filter.
The removed sentenced were in German, French, Spanish, and Dutch as well as some parse errors, e.\,g., in equations, citations and abbreviations.}
We use a simple regular expression (\texttt{\textbackslash .\textbackslash s?[A-Z]}) 
for sentence splitting
over readily-available NLP libraries as the latter tend to cut statistics in the middle of a sentence, since they include a ``.''.
\extended{Thus, patterns we are interested in like a statistic reported in APA style is susceptible to be cut by the state of the art methods.} 
An example of a typical sentence with statistics conform to APA style is: 
``\textit{The results of the paired sample $t$-tests indicated that negative emotion after inducement was significantly higher than at baseline (t(56) = 13.453, p < .05)}''.
Especially the last part of the statistical record \textit{[...]p < .05)} is susceptible to be cut.
Furthermore, if a sentence is split in the statistic record or somewhere else, it might make it impossible to determine the experiment conditions and experiment topics.
Each sentence not containing digits is filtered out\extended{~because they would not contain any statistics}.

\extended{Furthermore, through the process of converting PDF or HTML files to JSON, some conversion errors may occur.
One such sentence found in CORD-19 is (from ~\cite{duizer}): ``\textit{Inactivation at $100 \propto C$ was, however, complete within seconds (Duizer et al., 2004a) .The resistance of FeCV (in suspension) to inactivation by UV 253.7 nm radiation was reported to be highly variable.}''
The first observation made is, that $100 \propto C$ should be $100$ °$C$.
These kind of errors can also occur in statistics. 
This will make the process of identifying a pattern much harder, because, in this case, it is not to be expected to find a temperature notation written like this. 
Second, this is not one single sentence, it is actually two different sentences, but due to the lack of a white-space character after the "." of the first sentence, the splitting sequence did not detect the end of the sentence and therefor interprets these two sentences as one.

This instance can be intercepted by slightly modifying the splitting pattern to make the white-space "\texttt{\textbackslash s}" optional, like "\texttt{\textbackslash s?}".
However, it is possible, that some sentences can start with a digit or lower case character instead of an upper case character. 
Defining a pattern that will match all these cases could also lead to increased false positive rate and splitting in the middle of a sentence, corrupting the results. 
Thus, we did not apply it.}

\subsection{Wrapper Induction for Statistics Extraction}
\label{sec:activeWrap}
To extract the statistics of the preprocessed paper, a wrapper induction approach was applied to determine general rules to detect reported statistics and extract the type and values of this statistic.
Unlike existing wrapper induction approaches like~\cite{liu2007web} that operate on HTML document as input, the specific challenge we face is that the input sentences consist of plain, mostly unstructured text. 
Thus, no landmark tokens can be easily identified.
Furthermore, we cannot make any assumption about the number of statistics that are reported in a sentence.
There may be none, a single or in some cases even multiple statistics reported in a single sentence.
Finally, a sentence that contains digits and/or parentheses that are indicative for a statistic record may also be a false positive, which has to be filtered out. 

In order to address these challenges, we developed an approach based on two sets of rules, $R^+$ and $R^-$. 
The set $R^+$ resembles rules that actually refer to statistics reported in a sentence. 
$R^-$ is the rule set that confirms that a sentence does not contain statistics.
The $R^+$ rules support common types of inferential statistics, namely Pearson’s Correlation, Pearson Spearman Correlation, Student's $t$-test, ANOVA, Mann-Whitney U Test, Wilcoxon Signed-Rank Test, and Chi-Square Test.
But the concept is transferable to arbitrary types. 
Statistics whose type is not identifiable, \eg due to missing details in the reporting, are summarized under the  type ``other''. 
Sub-rules $S_i = \{s_{1}, \dots, s_{k}\}$ are defined for each statistics rule $r_i$ in $R^+$.
Thus, the elements of $R^+$ are actually tuples of the form $(r_i, S_i)$.
The rules $r_i$ are used to detect the different statistic types, such as a student $t$-test or Analysis of Variance (ANOVA). 
The rules $s_{j} \in S_i$ are used to detect the different statistic parameters. 
\extended{For example, in ``\textit{[...] (t(29) = -1.85, p = .074) [...]}'' the degree of freedom is $29$, the $p$-value is $0.074$, and the $t$-statistic is $-1.85$.}

The $R^+$ (together with its sub-rules) and $R^-$ rules are learned by wrapper induction.
The main loop of the learning process can be seen in Algorithm~\ref{alg:ActiveWrapperLoop}.
The algorithm can be applied to a whole corpus of documents, \ie it includes step 0.
It splits the documents into sentences (line~\ref{alg:split}), which are processed based on whether they contain any digits and whether these digits are already considered, \ie covered by a rule (see line~\ref{alg:covered}).
The respective statistic type, if detected in a sentence, is defined by the rule set $R^+$. 
If $R^+$ classifies a sentence as statistic with rule $r_i$, the respective sub-rules set $S_i$ are applied to extract the details (line~\ref{sec:details}).
If neither $R^+$ nor $R^-$ classifies a sentence, it is shown to the user (line~\ref{sec:user}). 
The user then adds a new rule to the respective rule set.

\begin{algorithm}[h]
\small
    \caption{Wrapper induction for extracting statistics}
    \begin{algorithmic}[1]
        \State Input: $D$  ~ // Document(s) to be processed
        \State Input: $R^+$ // Set of positive extraction rules (with sub-rules)
        \State Input: $R^-$ // Set of negative rules
        \State Output: $L$  // Statistics records extracted from $D$
        
        \State $L \gets \emptyset$ ~~~ // Initialize empty output list 
        \State \label{alg:split} $S \gets$ $D$\text{.split} // Split $D$ into a set of sentences 
        \While{$S \ne \emptyset$}
            \State $s \gets$ $S$\text{.nextElement()} // Process next sentence string
            \State // String-based processing of each $s$        
            \While{$s$ contains unclassified numbers} \label{alg:covered} // String not empty
            \State $\text{statsType}$, $\text{subR}^+$ $\gets \text{apply}(R^+,s)$         

            \If{$statsType \neq \text{NONE}$}  
            \State // Found a stats using $R^+$, so extract values
            \State $\text{statsRecord}, s$ $\gets$ apply($subR^+$,s) \label{sec:details} 
            \State $L\text{.add(statsRecord)}$ // ... and add to output
            \Else
               \State // no stat found? get confirmation from $R^-$ rule
               \State $\text{nonStat} \gets \text{apply}(R^-,s)$ // Confirmation successful?
               \If{$\text{nonStat} = \text{FALSE}$} 
                \State // If neither $R^+$ nor $R^-$ work on string ...
                \State  $\text{Invoke}( \text{``'ask user' to add rule for s''})$ \label{sec:user}
               \EndIf
            \EndIf
               
            \EndWhile
        \EndWhile
    \end{algorithmic}
    \extended{\begin{flushleft}
    This algorithm describes the procedure to learn the rule sets $R^+$ and $R^-$ including the related subrules for statistic extraction. 
    \end{flushleft}}
   \label{alg:ActiveWrapperLoop}
\end{algorithm}

Consider the following example sentence to illustrate the wrapper induction approach:
``\textit{The independent sample t-tests indicated that there were not significant differences in the effect of ibuprofen 400 between males and females, (t(29) = -1.85, p = .074).}''
First the $R^+$ rules will be applied. Therefore, the \textit{t-test} match is found first by a rule like this regular expression: 
\texttt{(\textbf{?P<ttest>}\textbackslash (t\textbackslash s?\textbackslash (\textbackslash d+\textbackslash )\textbackslash s?=\textbackslash s?\textbackslash d+\textbackslash .\textbackslash d+ \textbackslash, \textbackslash s?[p,P] \textbackslash s? <?=? \textbackslash s? \textbackslash d+\textbackslash .\textbackslash d+ \textbackslash ))}.
The part \texttt{?P<ttest>} defines the type of the statistic, here a Student's $t$-test.
The match of the rule in the sentence is \textit{(t(29) = -1.85, p = .074)}. 
Out of this sub-sentence, the detailed values of the $t$-test will be extracted by using the respective sub-rules.
The sub-rules have the same structure as the main $R^+$ rule, except that the different statistic parameters are tagged.
For example, \texttt{P<pval>} means that the following rule extracts the $p$-value:
\texttt{t\textbackslash s?\textbackslash (\textbackslash d+\textbackslash )\textbackslash s?=\textbackslash s?-?\textbackslash d+\textbackslash.\textbackslash d+\textbackslash ,}
\texttt{\textbackslash s?[p,P]\textbackslash 
s?<?=?\textbackslash s?}
\texttt{
(\textbf{?P<pval>}\textbackslash d+ \textbackslash . \textbackslash d+)}.
The extracted values from the example above are $df = 29$, $t = -1.85$, $p = 0.074$ together with the sentence fragment. 
But the sentence still contains digits in ``\textit{ibuprofen 400}''. 
When there is no $R^+$ rule left (like for this case), a corresponding $R^-$ rule should match the $400$, e.\,g.:
\texttt{[a-zA-Z]+\textbackslash s\textbackslash d+ \textbackslash s[a-zA-Z]+}.
If there are no digits left in the sentence, which are not covered by some rule, either $R^+$ or $R^-$, the sentence is completed. 
The next sentence is processed, until there are no more sentences left.
Result of step 1 is a set of sentences, known to contain statistics.
These are further analyzed to extract the experiment conditions and experiment topics.

\subsection{Wrapper Induction for Condition Extraction}
\label{sec:ruleBased}
\label{sec:gbce}

We apply a second wrapper to extract experiment conditions from the statistic sentences provided by step 1.
This is motivated from the assumption that the input sentences should report, besides the statistic, also the experiment conditions of the statistics.
To extract the conditions, we use the off-the-shelf tool spaCy for part-of-speech tagging and extracting grammatical dependency trees.\extended{~While generally a trained statistical entity recognition model would be a preferred approach\footnote{\url{https://spacy.io/usage/rule-based-matching}}, we follow a rule-based approach for condition extraction due to the lack of training data.}

The principle idea of our grammar-based condition extraction (GBCE) approach is the detection of nominal phrases in the sentences provided by step 1.
Thus, the idea behind GBCE is that all information about experiment conditions contains such a noun.
A nominal phrase is a syntactic, self-contained unit, whose core consists of a noun.
All phrases that are not part of a noun phrase (or dependent components of the noun phrase) can be ignored by rules when extracting the experiment conditions.
For example, in the sentence ``\textit{There is a positive statistically significant correlation between perceived knowledge and measured basic knowledge}'' a noun phrase would be ``\textit{a positive statistically significant correlation}'' with the core ``\textit{correlation}'', whereas adjective and article are dependent companions of the nominal head. 

Using the spaCy library, we annotate the sentences with linguistic knowledge such as UnifiedPOS (UPOS) tags, the 
extended POS tags, e.\,g., that the verb is past tense or the noun is proper singular, and the syntactic dependency describing the relation between tokens, e.\,g., preposition and object of preposition. 
This forms a parse tree with grammatical annotations and dependencies between tokens, which are used for rule-based matching.
\extended{An example parse tree is shown in Figure~\ref{fig:spacy}.}
\extended{The data structure of the parse tree that spaCy is working on is a so called \texttt{Doc} object\footnote{\url{https://spacy.io/api/doc}}.
It is a container of a sequence of tokens for accessing linguistic annotations.
After tokenizing every word of the input sentence, the \texttt{Doc} object is processed.
The default pipeline consists of a tagger, where each token is assigned a POS tag, a parser, adding dependency labels to aid natural language processing, and an entity recognizer.
This makes it possible to put the individual words of a sentence into context and thus forms a tree structure, which we refer in the follow as \textit{grammar tree}.
To customize the default variant, it is possible to exclude default methods and add custom methods. }
SpaCy allows to iterate through the sentence with the part-of-speech annotations and extracting grammatical dependencies in two ways.
On the one hand, the parse tree can be processed in sequential token order. 
This was mainly used for creating condition extraction rules.
On the other hand, the sentence can also be navigated following the parse tree. 
This was mainly used for extracting noun phrases.
\extended{\begin{figure*}
    \centering
    \includegraphics[width=0.8\textwidth]{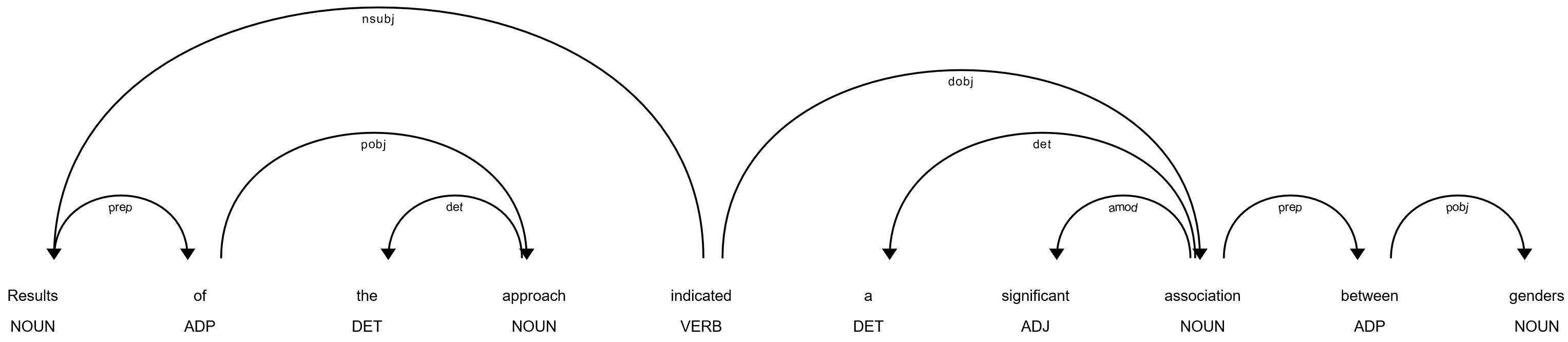}
    \caption{Example of a parse tree sentence with experiment conditions, image created with spaCy visualization module.}
    \label{fig:spacy}
\end{figure*}}
Further processing for grammar-based condition extraction is needed.
This includes removing all content within parentheses, as they interfere with the dependency parser and noun phrases could be detected incorrectly. 
It is safe to remove the content of the parentheses, since it  contains the statistics that is already extracted in step 1.
Subsequently, noun phrases are being identified, including their associated grammatical modifiers. 
\extended{We consider the following modifiers: numeric-, prepositional-, adverbial-, nominal-, appositional-, adjectival, adverbial clause-modifier and clausal modifier of nouns (adjectival clause).}
\extended{The tool spaCy is not capable of correctly processing quotation marks. 
To address this problem, if a noun phrase was identified inside quotation marks, the whole quotation gets included into the noun phrase. 
Also, spaCy interprets the usage of a semicolon as a new sentence.
This would result in two separate parse trees, which our rules are not designed for.
Since no experiment conditions were found after a semicolon, the rest of the sentence was excluded.}

\label{par:ruleset}
After preprocessing the input data, rules are learned with the goal of extracting experiment conditions based on noun phrases. 
Similar to the wrapper induction for statistic extraction (see Section~\ref{sec:activeWrap}), the GBCE operates with two rule sets $R^+$ and $R^-$, since the functionality is analogously. 
The difference is that instead of classifying numbers as statistic candidates they are confirming tokens as noun phrases or removing them.
If a noun phrase can be classified as experiment condition by a $R^+$ rule, the information gets extracted and the noun phrase of that sentence will not be considered further. 
In general, if no noun phrases are left to assign or there are no $R^+$ or $R^-$ rules left to be applied, the wrapper stops and outputs the results. 

When learning the rules through the wrapper, it was possible to determine specific grammatical patterns that never included experiment conditions and thus were added to the $R^-$ rule set.
The $R^-$ rules includes patterns such as personal pronouns, \eg ``\textit{we found}''.
Another $R^-$ rule excludes aspects, which is the case when the root of the sentence is not the main verb but instead a passive auxiliary.
\extended{A passive auxiliary is a subclass of verbs that add functional or grammatical meaning to the main verb.}
In terms of $R^+$ rules, there are rules that, when matched, all experiment conditions can be extracted and no further rules need to be applied. 
These rules exploit the fact, that the English grammar often follows specific patterns.
\extended{One such pattern that can be exploited is that the sentences often include comparative adjectives when describing the experiment conditions.
Those are mainly used to compare differences between two objects.}
An example pattern is: ``\textit{Noun (subject) + verb + comparative adjective + than + noun (object)}''.
If a perfect match is not possible, a sub-rule set is applied for locating the experiment conditions.
An example is the rule that is identifying relative clauses, introduced by interrogative words.
\extended{Relative clauses are non-essential parts of a sentence.
They only add additional meaning to a noun phrase.}
If a relative clause is identified, it gets included into the noun phrase.
When the relative clause is started by an interrogative word, the noun phrase is an experiment condition.
For example: 
``\textit{Par\-ti\-ci\-pants who agreed that the COVID-19 outbreak was threatening their livelihood...}''.
Another $R^+$ rule for extracting experiment conditions is checking for enumerations.
This rule recognizes an enumeration, splits the elements, and stores them as separately conditions.

\subsection{Aspect Model for Topic Extraction}
\label{sec:ABAE}
The final component of STEREO extracts the general topic of the sentence.
Here, we adapt an unsupervised algorithm for aspect extraction (ABEA)~\cite{abae}, which combines word embeddings into a sentence embedding via an attention mechanism and then compresses the information further with an autoencoder-like structure to create an aspect embedding. 
Like GBCE, the input to the adapted ABAE approach are the sentences extracted from step 1 containing statistics.
For topic extraction, we assume that there are $K$ different aspects in the documents of the CORD-19 corpus, \ie $K$ different experiment topics that in principle can be described by the sentences.
The aim is to identify per sentence, the \textit{specific} instance of an aspect that the sentence can be classified to. 
Thus, the number aspects $K$ can be generally quite high as there can be many different experiment contexts described in the sentences.
For instance, the aspect extracted as experiment topic from the example sentence in the introduction is ``personal data''.
In contrast, the number of aspects $K$ considered in the original ABAE paper were rather small, because there is only a limited number of relevant aspects in reviews over objects such as restaurants.
\extended{For example, $K=14$ was used for aspects extracted from restaurant reviews such as food, service, price, etc.}
Since we cannot make such an assumption, we trained ABAE with different values of $K$.
We evaluate which model delivers better results.

As ABAE is unsupervised, its aim is to maximize the difference between embedding of the input sentence and the average word embedding of any negative sample.
A negative sample is a sentence from the input data with a different aspect than the current sentence.
As ABAE is unsupervised, neither the aspect of the current sentence nor the aspect of any other sentence is known before and during training. 
The negative samples are randomly drawn from the input data for each input sentence\extended{~and over multiple training epochs}. 
The sentence embeddings are combined with an aspect embedding matrix, which is optimized during learning to improve diversity of the aspects.
Finally, the most representative words of each aspect are extracted from the word and aspect embeddings and the aspects are manually inferred from those.
\extended{A detailed description of ABEA and its training and evaluation is provided in the supplementary material, see Section~\ref{sec:detailsofabea}.}

\section{Experimental Apparatus}
\label{sec:expApp}

\extended{We evaluate each step of our pipeline separately.
The dataset and procedures are described in the following subsections.}

\subsection{Dataset}
We use the Covid-19 Open Research Dataset (CORD-19).\footnote{\url{https://kaggle.com/allen-institute-for-ai/CORD-19-research-challenge}}
\extended{It has been constructed to enable a basis to develop text and data mining tools that help the medical community answer high priority scientific questions, especially regarding the COVID-19 pandemic.
This dataset has been created by a coalition of the White House and leading resource groups.}
It includes around $200,000$ scientific articles (21st September 2020) of which over $108,000$ are scientific full text papers about COVID-19, SARS-CoV-2, and related corona viruses.
We pre-processed the dataset as described in Section~\ref{sec:preprocessing}, resulting in $16,141,291$ sentences.
We identified how many sentences potentially contain a test statistic, \ie how many contain at least a single digit.
From all sentences in CORD-19, about $55$\% contain at least a single digit.
These were used as input for our approach.

\subsection{Procedure and Evaluation Measures for Statistics Extraction}
To learn the $R^+$ and $R^-$ rule sets, we applied the wrapper induction approach from Section~\ref{sec:activeWrap} on the CORD-19 dataset.
\extended{Documents are processed in the order of how they are organized in the dataset, which is according to a random index.}
We trained the wrapper on the sentences of the first $500$ documents and analyzed the results.
To evaluate the statistic extraction, we took a random sample of $200$ non-statistic and statistic sentences, except when there where less.
In that case, we took all found sentences.
We did this for each type of statistic, once for sentences in APA conform writing style and for non-APA conform reports.
\extended{We regard a rule conform to APA style, if all parameters are present and the formatting is correct.
However, we tolerate little derivations from the APA style formatting, i.\,e., $P=0.07$ is not conform to APA, because the $P$ is a capital letter and there is a leading zero before the ``.''.}
An APA-conform sentence was classified as correct, if all attributes of the statistic could be extracted.
The non-APA conform sentences were classified as correct, if the type of statistic was correctly detected.
The sentences were extracted by the learned $R^+$ and $R^-$ rules.

We manually determine the true positives (tp), true negatives (tn), false negatives (fn), and false positives (fp).
Two assessors were responsible for this classification.
In ambiguous cases, a consensus was found.
As measures we use precision $prec = \frac{tp}{tp + fp}$ and a count of how many sentences were extracted in total.
\extended{We used the combination of $prec$ and amount of extracted samples, because for some statistic types, we could only obtain a small amount of samples.
For these samples, the precision might not be expressive.}
Furthermore, a coverage of our $R^+$ and $R^-$ rules was calculated by taking a random sample of $10,000$ unseen documents from CORD-19\extended{~and determining the proportion of covered sentences in these documents}.

\subsection{Procedure and Evaluation Measures for Condition Extraction}
\label{gbceProcedure}

The rules for GBCE were learned on a sample of $130$ sentences provided by the statistic extraction from step 1 of the pipeline (see Section~\ref{sec:activeWrap}).
To learn the $R^+$ and $R^-$ rule sets, we applied the wrapper induction approach and went through the grammar trees of each sentence while manually checking if the experiment conditions were extracted correctly. 
If this was not the case, an already existing rule was adapted, a new rule was created, or specific words that often occurred were added to the bag-of-words.

For evaluation of the GBCE, we randomly sampled $200$ sentences from the set of sentences provided by the statistic extraction in step 1.
The sentences were evenly sampled to form a set of $100$ sentenced being conform to APA writing style and $100$ sentences that are not non-APA conform.
We manually checked the output of the grammar-based condition extraction rules if they correctly identified noun phrases as the experiment conditions.
This was done by agreement of two different reviewers.
If their evaluation differed, it was discussed and an agreement reached.

\subsection{Procedure and Evaluation Measures for Topic Extraction}
\label{sec:ProcABAE}

Multiple ABAE models were trained on different embeddings, subsets of the CORD-19 data sets, and with different numbers of aspects $K$.
\extended{We explain the parameter choices for the different models, their training, as well as how the models were evaluated.}
Three different subsets of CORD-19 were used to train and evaluate the topic extraction with ABEA.
These subsets are first, \textbf{cord}: all sentences from the preprocessed CORD-19 dataset as described in Section~\ref{sec:preprocessing}.
Second, \textbf{all-sen}: extracted sentences containing \textit{any} statistics.
Third, \textbf{supp-sen}: extracted sentences containing only the following statistics: Student's $t$-test, Pearson Correlation, Spearman Correlation, ANOVA, Mann-Whitney U, Wilcoxon Signed-Rank, Chi-Square.
Stopword removal and lemmatization was applied to all three datasets.
\extended{The number of unique and total words as well as the number of sentences of each dataset are shown in Table~\ref{tab:embSizes}.

\begin{table}[h]
    \centering
    \begin{tabular}{|l|r|r|r|}
        \hline
        Dataset &  unique words &  total words & sentences\\
        \hline
        cord & $1,400,093$ & $238,582,456$ & $16,141,291$\\
        all-sen & $45,071$ & $1,467,485$ & $113,147$\\
        supp-sen & $7,189$ & $81,936$ & $6,092$\\
        \hline
    \end{tabular}
    \caption{Number of unique words, total number of words, and sentences per datasets, used for topic extraction with ABEA.}
    \label{tab:embSizes}
\end{table}
}

Three sets of Word2Vec (W2V)~\cite{word2vec} embeddings with dimension $d=200$ were trained, one each on cord, supp-sen, and all-sen. 
We used the skip-gram algorithm and a window size and negative sampling of $5$.
\extended{For the cord embedding, the number of words in the embeddings was limited to about $50,000$ by choosing $100$ as the minimum word frequency.
This was done because most of the infrequent words do not contain information that the embedding can learn as well as to reduce the model size.
The most frequent $50,000$ words cover about $97\%$ of the total words in the cord dataset.
The distribution of word occurrences covered by the number of unique words is plotted in Figure~\ref{fig:wf}.
The other two datasets contained less than $50,000$ words.
Thus, for all-sen the minimum word frequency was left on W2V's standard of $5$, which includes about $96.4\%$ of the total words.
For supp-sen, the minimum word frequency was chosen to be $3$, which covered $94.1\%$ of the total words.
This is a trade-off between giving W2V enough examples for every word included in the embeddings and covering enough words of the dataset to have a meaningful result after applying the embedding. 
For example, for supp-sen, the W2V standard setting of $5$ would have covered only $88.9\%$ of the total words.
\begin{figure}
    \centering
    \includegraphics[width= 0.48\textwidth]{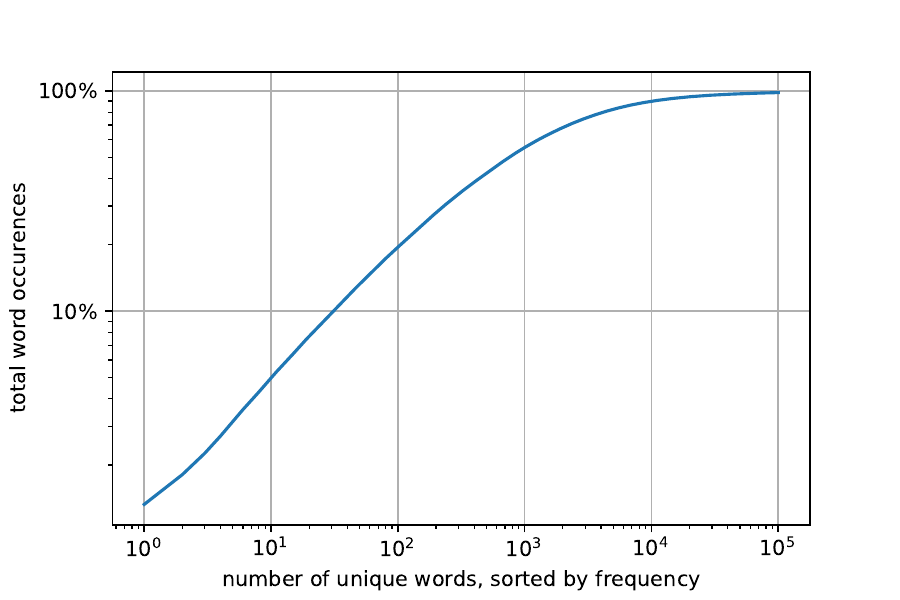}
    \caption{Total word coverage over number of unique words to $100k$ words on the CORD-19 dataset after pre-processing. 
    We chose a cutoff of $50k$ as that covers about $97.0\%$ of the total words occurrences and there is very little additional return for higher values, e.\,g. $98.2\%$ of the words are covered at $100k$.}
    \label{fig:wf}
\end{figure}}
After training the W2V embeddings, we trained the ABEA models.
We chose to limit the longest supported sentence length be $70$ words, as this covers over 99\% of all sentences in the dataset.
All shorter sentences got padded to that length.
\extended{The relation between the longest sentence length and coverage of the CORD-19 dataset can be seen in Figure~\ref{fig:sl}.}
The number of negative samples $m$ was set to $20$.
Different values for the number of aspects $K$ were tested, namely $15$, $30$, and $60$.
Thus, different ABEA models were trained once on each dataset with one of the three sets of W2C embeddings and using the three different values for the number of aspects $K$.
The only exception was the combination of the W2V embeddings trained on supp-sen with the ABEA model trained on all-sen, as the training data would contain many unknown words for the embedding.
The weight of the regularization in the loss function was set to $\lambda = 1$ like in the original paper~\cite{abae}.

\extended{\begin{figure}
    \centering
    \includegraphics[width= 0.48\textwidth]{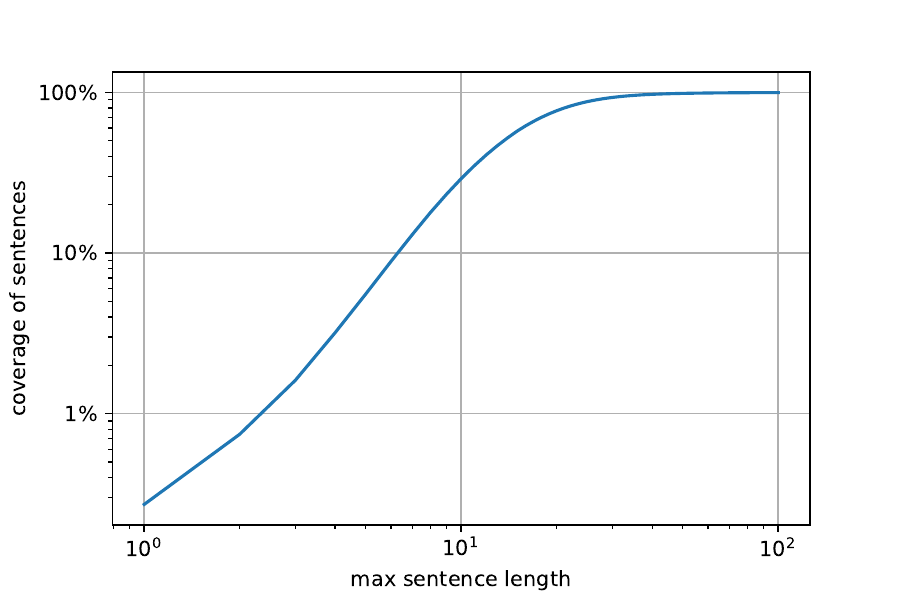}
    \caption{Coverage percentage over sentence lengths to $100$ on cord dataset after preprocessing. We chose a cutoff of $70$ as that covers $99.5\%$ of sentences. Most of the very long sentences contain parse or sentence splitting errors.}
    \label{fig:sl}
\end{figure}}

\extended{The embedding matrix $E$ got initialized with one of three the pre-trained W2V embeddings and was then fixed during training of the other parameters of ABAE.
The aspect embedding matrix $T$ got initialized with the centroids of a $k$-means run on the word embeddings.
The matrices $M$, $W$, and the vector $b$ were initialized randomly.
$M$, $W$, $b$ and $T$ were trained with Adaptive moment estimation (Adam).
Adam was used for $50$ epochs with a learning rate of $0.01$, epsilon of $10^{-7}$, a batch size of $10$ on the smaller and $32$ on the larger dataset, and all other parameters left to the standard setting.
Like done in the ABAE paper the model after the epoch with the smallest loss was saved and evaluated.}

After training, the aspects, \ie the experiment topics were inferred manually from the set most representative words.
\extended{See the Appendix~\ref{sec:detailsofabea} for details.}
If the representative words did not contain any concise groups of words, the aspect was set to ``miscellaneous'', which will always be evaluated as wrong.
For evaluation of topics extracted with ABEA, the same randomly sampled $200$ sentences were used as for the evaluation of GBCE. 
The sentences were sampled such that $100$ were APA conform and $100$ non-APA conform sentences.
All ABEA models for topic extraction were evaluated on the same $200$ sentences by manually checking if the model extracted a correct aspect.
This was done by agreement of two different reviewers, if their evaluation differed it was discussed to reach agreement.

\section{Results}
\label{sec:results}

\extended{This section presents the results of our experiments.
The first subsection is about the statistic extraction with the rules learned by the wrapper induction approach.
The second subsection covers the conditions extraction and the third the experiment topics extraction.}

\subsection{Results for Statistics Extraction}
We applied the wrapper induction approach for rules introduced in Section~\ref{sec:activeWrap} on the first $500$ documents of the CORD-19 dataset, which contained a total of $38,099$ sentences.
The result is a set of $85$ $R^+$ rules extracting statistics, and a set of $1,425$ $R^-$ rules, which classify some digits as non-statistic.
We checked the coverage of the rules over a random sample of $10,000$ unseen documents in the CORD-19 dataset.
It showed that the rules learned on $500$ documents, \ie $0.25\%$ of the corpus, cover $95\%$ of the sentences in the sample.

In Table~\ref{tab:sizes}, one can see how many sentences containing statistics were extracted from the \textit{whole} CORD-19 dataset.
The results are shown per statistic type and based on whether the reported statistics was conform to APA style or not.
Note, we focused learning rules on the common inferential statistics used in life sciences, psychology, etc.
Other statistics such as odds ratio, interquartile range (IQR), etc., are subsumed under ``other statistics''.
The row ``non determinable'' refers to cases where only a $p$-value was reported, \ie it was clear that this is an inferential statistic, but because of lack of further information the type of statistic could not be determined.
As can be seen from the table, over $113$k reported statistics could be extracted, of which $<1\%$ is APA conform.

\begin{table}[h]
\small
    \centering
    \begin{tabular}{|l|r|r|}
        \hline
        Statistic type &  APA &  non-APA \\
        \hline
        Student's $t$-test & 608 & 179\\
        Pearson Correlation & 113 & 4,962\\
        Spearman Correlation & 1 & 528\\
        ANOVA & 0 & 9\\
        Mann-Whitney U & 2 & 34\\
        Wilcoxon Signed-Rank & 0 & 0 \\
        Chi-Square & 14 & 31\\
        \hline
        Other statistics & not applied & 19,151\\
        Not determinable (only $p$-value) & not applicable & 87,904\\
        \hline
        \textit{Total number of extracted statistics}  & \textit{738} & \textit{112,798} \\
        \hline
    \end{tabular}
    \caption{Number of reported statistics from the \textit{whole} CORD-19 dataset, for APA and non-APA conform statistics. 
    ``Other statistics'' are, \eg odds ratio, IQR etc.
    If only a $p$-value was reported, the type of statistic is ``not determinable''.}
    \label{tab:sizes}
\end{table}

We manually evaluated the quality of our $R^+$ extraction results over $200$ random samples per statistic type and split by APA conform and non-APA.
Thus, in total we evaluated $1,383$ reported statistics, $330$ APA conform and $1,053$ non-APA conform.
The precision values for each statistic type are shown in Table~\ref{tab:precision}.
We achieve a precision of $100\%$ for all APA conform statistics (ANOVA and Wilcoxon Signed-Rank did not occur in APA conform writing style).
In the case of non-APA conform report, the precision ranged from $91\%$ to $100\%$.
The Wilcoxon Signed-Rang test could not be evaluated, since we did not extract any sentence reporting that statistic.
\extended{The smallest precision for statistic extraction was for the non-APA Student's $t$-test with $91\%$.}

\begin{table}[h]
\small
    \centering
    \begin{tabular}{|l|r|r|}
        \hline
        Student's $t$-test   & $100\%$ & $91\%$ \\
        Pearson Correlation  & $100\%$ & $98\%$ \\
        Spearman Correlation & $100\%$ & $100\%$ \\
        ANOVA                & n/a     & $100\%$ \\
        Mann-Whitney U       & $100\%$ & $100\%$ \\
        Wilcoxon Signed-Rank & n/a  & n/a \\
        Chi-Square           & $100\%$ & $97\%$ \\
        \hline
        Other statistics     & - & $95\%$ \\
        \hline
    \end{tabular}
    \caption{Precision values for the extraction of reported statistic.
    The precision is calculated on $200$ samples for each statistic type and for both APA conform and non-APA reporting. 
    \extended{If less than $200$ samples were available for some statistic type, the precision has been calculated only the respective amount of extracted samples.}}
    \label{tab:precision}
\end{table}

The amount of sentences covered by our $R^+$ rules was $95$\%.
Thus, we also checked a random sample of $200$ sentences from the uncovered sentences, if they contained statistics we have not learned.
Of this sample, $21$ contained some statistic, $157$ were without statistic. 
$22$ of the sentences contained a text conversion error in the CORD-19 dataset, independent of them containing statistics or not.
An example for such an error is the sentence ``\textit{Notably, however, CD8a - DCs and also pDCs can cross-prime CD8 + T-cell responses under certain conditions (102) (103) (104) 123)}''.
However, the original sentence is ``\textit{[...] responses under certain conditions (102–104, 123)}''\footnote{\url{https://www.ncbi.nlm.nih.gov/pmc/articles/PMC4603245/}}.
To evaluate the $R^-$ sentences, \ie to determine if the negative rules successfully rejected numbers in a sentence as non-statistics, we took another random sample of $200$ sentences.
This sample is taken from the $R^-$ matches on the CORD-19 corpus, except the first $500$ documents we used for training.
Of this sample\extended{~of $200$ sentences}, $99.5 \%$ were correctly classified as not containing a statistic report.

Regarding the non-APA conform statistics we extracted, it is interesting to understand what specific statistical parameter was missing in the report, \eg the degree of freedom, etc., and how many times this parameter was missing in the sample.
Tables~\ref{tab:MissParaTtest} to~\ref{tab:MissParaCS} report this information per statistic.
The diagonal show how many times a parameter was missing on its own.
In the other entries, one can see how often a pair of parameters was missing.
For example, in Table~\ref{tab:MissParaTtest} the row degree of freedom (doF) and column $t$-value (tval) shows how often doF and tval were missing together.
The column margin shows how often a value was missing, either alone or in combination with another parameter. 
One can see that the elements in the diagonal are all $0$, \ie no parameter was missing on its own.
In contrast, in Table~\ref{tab:MissParaSpearman} one can see that doF was the only missing parameter in $527$ samples.

\begin{table}
   \small
    \centering
    \begin{tabular}{|l|r|r|r||r||r|}
        \hline
        Missing Parameter&  doF & tval & pval & other & Sum\\
        \hline
        Degree of freedom (doF) & 0 & 75 & 21 & 1 & 97\\
        \hline
        $t$ value (tval) & \cellcolor{lightgray} & 0 & 0 & 1 & 76\\
        \hline
        Significance level (pval) & \cellcolor{lightgray} & \cellcolor{lightgray} & 0 & 1 & 22\\
        \hline
    \end{tabular}
    \caption{Count of missing parameters from non-APA conform Student's $t$-tests samples. Calculated on $179$ samples\extended{~(\ie all extracted non-APA conform $t$-tests)}. 
    Column ``other'' refers to samples where all parameters doF + tval + pval were missing.}
    \label{tab:MissParaTtest}
    \small
    \centering
    \begin{tabular}{|l|r|r|r|r|r|r|}
        \hline
        Missing Parameter &  doF & rs & pval  & Sum \\
        \hline
        Degree of freedom (doF) & 527 & 0 & 1 & 528\\
        \hline
        Spearman correlation (rs) & \cellcolor{lightgray} & 0 & 0 & 0\\
        \hline
        Significance level (pval) & \cellcolor{lightgray} & \cellcolor{lightgray} & 0 & 1\\
        \hline
    \end{tabular}
    \caption{Missing parameters from non-APA conform Spearman Correlation samples. 
    Calculated on $528$ samples.}
    \label{tab:MissParaSpearman}
    \small
    \centering
    \begin{tabular}{|l|r|r|r||r||r|}
        \hline
        Missing Parameter &  doF & r & pval & other & Sum \\
        \hline
        Degree of freedom (doF) & 4961 & 0 & 0 & 1 & 4962\\
        \hline
        Pearson Correlation (r) & \cellcolor{lightgray} & 0 & 0 & 1 & 1\\
        \hline
        Significance level (pval) & \cellcolor{lightgray} & \cellcolor{lightgray} & 0 & 1 & 1\\
        \hline
    \end{tabular}
    \caption{Missing parameters from non-APA conform Pearson Correlations. Calculated on $4,962$ samples. 
    Column ``other'' refers to samples with all parameters doF + pval + r missing.}
    \label{tab:MissParaPearson}
    \small
    \centering
    \begin{tabular}{|l|r|r|r|r||r||r|}
        \hline
        Missing Parameter &  doF & fval & pval & r & other & Sum \\
        \hline
        Degree of freedom (doF) & 0 & 0 & 0 & 0 & 2& 2 \\
        \hline
        F value (fval) & \cellcolor{lightgray} & 0 & 0 & 7 & 2 &9\\
        \hline
        Significance level (pval) & \cellcolor{lightgray} & \cellcolor{lightgray} & 0 & 0 & 1 &1\\
        \hline
        Effect size (r) & \cellcolor{lightgray} & \cellcolor{lightgray} & \cellcolor{lightgray} & 0 & 2 &9\\
        \hline
    \end{tabular}
    \caption{Missing parameters from non-APA conform ANOVA samples. Calculated on $9$ samples. 
    Column ``other'' refers to either doF + fval + r or doF + fval + r + pval missing.}
    \label{tab:MissParaAnova}
    \small
    \centering
    \begin{tabular}{|l|r|r|r|r||r||r|}
        \hline
       Missing parameter &  U & z & pval & r & oher & Sum \\
        \hline
        U & 4 & 0 & 0 & 0 & 13 & 17\\
        \hline
        z & \cellcolor{lightgray} & 16 & 0 & 0 & 13 & 29\\
        \hline
        Significance level (pval) & \cellcolor{lightgray} & \cellcolor{lightgray} & 0 & 0 & 0 & 0\\
        \hline
        Effect size (r) & \cellcolor{lightgray} & \cellcolor{lightgray} & \cellcolor{lightgray} & 0 & 13 & 13\\
        \hline
    \end{tabular}
    \caption{Missing parameters from non-APA conform Mann-Whitney U samples. Calculated on $34$ samples. 
    The column ``other'' refers to samples where U + z + r was missing.}
    \label{tab:MissParaMWU}
    \small
    \centering
    \begin{tabular}{|l| r| r |r |r||r||r|}
        \hline
        Missing Parameter &  $\chi^2$ & $N$ & pval & V & other & Sum\\
        \hline
        $\chi^2$-value ($\chi^2$) & 0 & 0 & 0 & 0 & 2 & 2\\
        \hline
        Observations ($N$) & \cellcolor{lightgray} & 0 & 0 & 1 & 2 & 3\\
        \hline
        Significance level (pval) & \cellcolor{lightgray} & \cellcolor{lightgray} & 0 & 0 & 2 & 2\\
        \hline
        $V$ &  \cellcolor{lightgray} & \cellcolor{lightgray} & \cellcolor{lightgray} & 28 & 2 & 31\\
        \hline
    \end{tabular}
    \caption{Missing parameters from non-APA conform Chi-Square samples. 
    Calculated on $31$ samples. 
    Column ``other'' refers to samples with $\chi^2$ + $N$ + pval + V missing.}
    \label{tab:MissParaCS}
\end{table}

\subsection{Results for Conditions Extraction}
For extracting the experiment conditions, we build the rules by manually analyzing $130$ sentences, which resulted in $35$ rules for GBCE.
Less sentences have been used for training GBCE's rules than for extracting statistics since the effort needed in defining grammar-based rules is much higher.
For learning grammar-based rules, we were able to discover a high amount of comparing adjectives as indicators for experiment conditions. 
This has prompted us to build several rules about this pattern. 
GBCE has been evaluated on $100$ samples that were APA conform and $100$ non-APA conform statistics. 
The results are shown in Table~\ref{tab:gbceRes}. 

\begin{table}
    \small
    \centering
    \begin{tabular}{|l|r|r|}
         \hline
         GBCE & APA & non-APA  \\
         \hline
         Correctly classified & 46 & 30\\
         \hline
         \hline
         Reason 1: Failed grammar & 4 & 5\\
         \hline
         Reason 2: Sentence structure & 10 & 3\\
         \hline
         Reason 3: Pre-processing error & 9 & 12\\
         \hline
         Reason 4: Dependency parser & 18 & 2\\
         \hline
         Reason 5: GBCE miss& 25 & 47\\
         \hline
    \end{tabular}
    \caption{Number of correctly extracted experiment conditions and reasons why the extraction failed. 
    In several samples, a combination of reasons were the cause.}
    \label{tab:gbceRes}
\end{table}

As one can see, the precision for extracting the correct conditions was with $46\%$ for APA conform statistics slightly higher than $30\%$ for non-APA conform statistics.
Every time the experiment conditions were not extracted correctly, we counted the number of occurrences and also identified the reason for its failure.
In several cases, there was a combination of different reasons.
Therefore, the sum of occurrences of reasons is higher than the amount of incorrect cases.
The reasons for failure are classified in five categories:
\textit{Reason 1: Failed to built a grammar-based representation}:  In these cases, false POS-tags or errors in the dependency parser of a sentence occurred.
This was mainly due to grammatical mistakes in the sentence structure or conversion errors in the COVID-19 dataset. 
\textit{Reason 2: Unusual sentence structure}: Describes errors due to unusual sentence structure such as a verb missing.
Since spaCy builds the parse tree with a verb as root, a missing verb affects the grammar-based process of finding conditions.
\textit{Reason 3: Error in pre-processing}:
As described in Section~\ref{sec:preprocessing}, the COVID-19 dataset has some errors in the pre-processing. 
Another case is when noun phrases were not correctly extracted.
\textit{Reason 4: Dependency parser wrongly splits sentence}
If a statistic was not correctly excluded form a sample before building the dependency parser, the statistic often misleads the dependency parser by splitting the sentence at the statistic.
\textit{Reason 5: GBCE misses the experiment conditions}:
Cases, where GBCE could not extract the conditions due to missing training for specific type of statistics or patterns.
All models were evaluated by both reviewers separately. 
The results were compared and, if different, an agreement reached.

\begin{table}
    \small
    \centering
    \begin{tabular}{|lll|r|r|}
         \hline
         Embedding & Training data & $K$ & APA  & non-APA  \\
         \hline
         supp-sen & supp-sen & 15 & 33 & 31\\
         \hline
         \textbf{supp-sen} & \textbf{supp-sen} & \textbf{30} & \textbf{75} & \textbf{73}\\
         \hline
         supp-sen & supp-sen & 60 & 28 & 28\\
         \hline
         \hline
         all-sen & supp-sen & 15 & 51 & 57\\
         \hline
         all-sen & supp-sen & 30 & 48 & 49\\
         \hline
         all-sen & supp-sen & 60 & 58 & 50\\
         \hline
         \hline
         all-sen & all-sen & 15 & 46 & 41\\
         \hline
         all-sen & all-sen & 30 & 7 & 22\\
         \hline
         all-sen & all-sen & 60 & 49 & 43\\
         \hline
         \hline
         cord & supp-sen & 15 & 38 & 37\\
         \hline
         cord & supp-sen & 30 & 42 & 44\\
         \hline
         cord & supp-sen & 60 & 62 & 54\\
         \hline
         \hline
         cord & all-sen & 15 & 57 & 56\\
         \hline
         cord & all-sen & 30 & 44 & 37\\
         \hline
         cord & all-sen & 60 & 33 & 46\\
         \hline
    \end{tabular}
    \caption{Number of correct ABAE outputs. Evaluated on $100$ samples in APA style and $100$ non-APA conform samples. 
    The first three columns shows which model was used: embedding dataset, training dataset, and number of aspects $K$.}
    \label{tab:abaeRes}
\end{table}

\subsection{Results for Topics Extraction}
We have five different combinations of embeddings and training data with ABEA for topic extraction as shown in Table~\ref{tab:abaeRes}.
For each combination, we trained three models for $K=15$, $30$, and $60$.
As explained in Section~\ref{sec:ProcABAE}, the embeddings are supp-sen, all-sen, and cord.
Training data are different subsets of CORD-19, namely sentenced that are supported by step 1 (supp-sen), sentences containing some statistics (all-sen), whether it was identifiable or supported, or not, and finally all CORD-19 sentences (cord).
The quality of the extracted topics was evaluated per model on $100$ sentences that strictly followed APA style and $100$ sentences with statistics not strictly following APA style. 
The number of correct topics can be seen in Table~\ref{tab:abaeRes}.
The model that performed best uses $K = 30$, an embedding trained on supp-sen, and the final ABEA model trained on supp-sen. 
As for GBCE, all ABEA models were evaluated by two reviewers. 
If classifications differed, consensus was reached.

\section{Discussion}
\label{sec:disc}

\subsection{Main Results}

The statistics extraction achieves a precision of $100\%$ for APA conform writing style. 
These results are in line with the precision of statcheck and its extension~\cite{Statcheck,DBLP:conf/www/LankaR0G21}.
In addition to APA conform patterns supported by statcheck~\cite{Statcheck} and its extension~\cite{DBLP:conf/www/LankaR0G21}, our pipeline can also extract non-APA conform writing styles for statistics.
The precision for extracting non-APA conform writing styles is $95 \%$, which is due to the high variety in which statistical analysis are reported.
From the $113$k statistics extracted from the entire CORD-19 dataset, over $99\%$ is not APA conform.
Thus, it is important to be able to extract non-APA statistics.

The basic idea of experiment condition extraction with GBCE is that sentences containing statistics mostly follow a common sentence structure.
Therefore, there should be a uniquely determinable finite set of rules that can exploit those patterns for extracting the experiment condition. 
However, deviations occurred frequently and we observe differences between statistical test types. 
For example, reporting conditions for correlations seem to follow a common pattern more often, while reporting conditions of chi-square tests did not.
Overall, we achieved a precision of $46\%$ for extracting conditions from APA-conform reports, which is notably higher than the $30\%$ for non-APA reporting.
The reason is that non-APA reporting generally has a higher variety. 
The sentences were longer, with a more complex structure, and also contained more statistics per sentence.
This is particularly evident from the higher number extraction failures due to  reason~$3$ (pre-processing error due to wrong sentences splitting) and reason~$5$ (GBCE misses condition due to variety in the reporting) as shown in Table~\ref{tab:gbceRes}.
\extended{The sentences containing multiple statistics were cropped during pre-processing.
This caused problems, if the context of the sentence was only comprehensible through neighboring sentences or if the sentence was not grammatically correct.}
One example is:
``\textit{The results show, that female participants used national newspapers [STATISTIC] highly significant less than male participants and international sources [STATISTIC] and YouTube [STATISTIC]}''.
In those cases, it was not possible to automatically distinguish between experiment topics and conditions.

Regarding the topic extraction from the reported statistic with ABAE, we found neither a pattern that models with lower or higher values of $K$ nor models with a specific embedding or trained on a specific dataset generally performed better. As the approach is unsupervised, we expected that the models perform similarly on the APA conform and non-APA conform sentences as shown in Table~\ref{tab:abaeRes}.
An overall high precision of over $70\%$ correctly extracted topics can be explained that most models have extracted at least one ``statistics'' aspect.
That is a correct result, but not useful for our use case, as that applies to every sentence that made it through step~1 of STEREO.
Overall, half of the correct answers were ``statistics''.
This can be addressed in a post-processing step, when the topic ``statistics'' can be filtered out to obtain the final list of extract topics.

\subsection{Threats to Validity}

For some statistic types like Pearson correlation and Student's $t$-test, we could extract many samples. 
For statistic types like ANOVA, where we just extracted $9$ samples, the results could be not representative enough. 
Nevertheless, the metrics for ANOVA are similar to the statistic types with more samples.
Therefore, it is plausible to assume that the results transfer to the types with few samples, too.\extended{\newline}
One may be surprised that we were not able to extract many statistics of every type.
Especially, for the Wilcoxon signed rank test, we did not observe a single occurrence (see Table~\ref{tab:sizes}). 
To rule out that this is purely due to fact that a Wilcoxon signed rank test was not part of the documents used for training, we manually added multiple extraction rules for Wilcoxon tests, which detect APA conform as well as some known non-APA deviations.
Using these rules, we could not extract a single example of a Wilcoxon signed rank test on the whole CORD-19 dataset.
This maybe due to the fact that there is none, or that Wilcoxon signed rank tests are written in such an unusual manner that we classified them as \textit{other}.
Likely there may be no Wilcoxon signed rank test as according to Weissgerber et al.~\cite{paramNonParam}, Wilcoxon tests are not commonly used.
\extended{In Weissgerber et al.~\cite{metaStat} they manually analyzed $328$ paper regarding their statistic writing style. 
They found many incomplete statistic reports in their sample from the PubMed dataset.
Since our dataset is also from the field of life science and we extracted a lot more non-APA conform statistics, we conclude that it is quite uncommon in life sciences to strictly follow this writing style.}

The rules for GBCE were created on the basis of the Collins Dictionary\footnote{https://grammar.collinsdictionary.com/grammar-pattern}\extended{~in cooperation with an anglistics student}.
\extended{We did not evaluate how the accuracy of GBCE differed between statistical test types.
Therefore, the results could deviate between different datasets with statistical test types that are not that common in the CORD-19 dataset.}
Since the evaluation procedure for GBCE did not vary from the evaluation of the aspect extraction, errors in GBCE evaluation did probably not occur as well. 
Especially, since the test cases were selected randomly, the overall results should generally fit to all possibilities.
In few cases, considering a single sentence only would be not sufficient for condition extraction. 
A future extension could consider the surrounding sentences, too.\extended{\newline}
Regarding topic extraction with ABAE, there are two steps where an error could occur. 
The first one is choosing the inferred aspect and the other one is evaluating whether a model found the right aspect for a sentence.
Nevertheless, all extracted terms and abbreviations were manually looked up, if needed.
Furthermore, the evaluation was done by consent of two assessors.

\extended{A completely different problem was the conversion from PDF to JSON in combination with different writing styles of statistical parameters.
For example, ``chi-square tests'' were written or got parsed as: chi-square, $\chi^2$, X\^{}2, or X2.
This is a commonly observed problem with any kind of metadata extraction tool on PDF to text converted documents such as scientific papers~\cite{DBLP:conf/www/LankaR0G21}.
To address this problem, we learned additional rules to include special characters introduced by the conversion process and different writing styles.}

\subsection{Generalization}

Our tool can be applied to datasets of different domains, \eg psychology, medicine, economics, etc., since APA is a common standard in different disciplines.
Thus, the rules should transfer to these domains, including the non-APA writing styles.
However, it would be beneficial to fine-tune the existing rule sets on a dataset from a new domain. 
Especially the $R^-$ rule set contains a lot domain-specific rules.
Additional rules for a new domain can be created by running the active wrapper process again.
\extended{The existing rules match already $95\%$ of the sentences in the CORD-19 dataset.
This is expected, because although there are different writing styles for statistics, it still follows a set of common rules, even if the writing is non-APA conform. 
Additionally, the extraction quality for $R^+$ and $R^-$ has been manually verified.
One may conclude that the same quality of extraction can be reached on other datasets.}\extended{\newline}
GBCE in general is a rule-based approach for experiment condition extraction. 
Since the different statistical test types and the English grammar including their structural patterns do not vary between domains, a generalization should be possible without further adjustments. 
\extended{The only restriction is that the language must be English, as a different language will have a different grammatical structure.}
Regarding the generalization of topic extraction, the embeddings and models could be reused if the dataset is in a similar domain. 
Otherwise, they could be either fine-tuned or retrained from scratch.
\extended{This means there have to be enough example sentences in the dataset to (re-)train the embedding and models.
If there is only a small number of example sentences, one could try a more general embedding, either like we did with the embedding trained on the whole CORD-19 dataset or a publicly available, most likely non-domain specific, embedding.
That means if there is enough data for training, in principal this approach can be applied to any similar problem.}

\extended{\subsection{Practical Impact and Future Work}}

\extended{Even for statistics perfectly written in APA style, typing errors can occur. 
Because of such errors, it is possible that our defined pattern for the specific statistic type might not find a match.
It might increase the amount of found statistics if the pattern is defined in such a way, that it can comprehend typing errors or even parse errors.
Thus, our tool is in general robust to these kind of errors.
This is a design feature to increase recall.
Additionally, we are explicitly able to distinguish the extracted statistic in APA and non-APA conform classification, which offers new possibilities for applications, e.\,g. reporting this back as feedback to the authors.}
\extended{On the other hand, generalizing such a pattern is not trivial, since tolerating typos could also lead to an increase in false positives.
To provide an aid in the learning process of the adaptive wrapper, one could implement an automatic test after creating a $R^-$ rule.
This new $R^-$ rule will be applied on a predefined $R^+$ sentence set, to check if the rule would classify the sentences correctly.
This way, the learn process would be supported to not overgeneralize and include false positives.}

\extended{An interesting direction of future work is to find a minimal set of extraction rules. 
During the process of wrapper induction, rules may arise that are (partially) covered by another rule or a subset of other rules. 
These rules could be removed in a post-processing step.
Since the equivalence of regular expressions is NP hard, this task is challenging and was yet not considered in the present work.} 

\extended{Regarding GBCE, the use of noun phrases as basis for reducing the point of interest was in general successful. 
However, one could think about adding further rule sets, specifically for the different types of tests.
On the one hand, this should achieve better results. 
On the other hand, the success depends on the re-usability of the rules that can be applied for every test.}
\extended{A risk is that this possibly needs a high number of specifically customized rules.
Since a purely rule-based approach performs well for structured patterns like IP addresses or URLs, it may be possible, that creating a labeled set of data for the training of a neural network could be a better trade-off than the creating rules for a rule-based approach. 
Especially since there are more differing structural patterns than expected.}

\extended{Regarding ABAE, there are some improvements that could be implemented for our tool.
More effort could be put into the dataset for tasks like sentence splitting, filtering of ``bad'' sentences, and to check and adapt the language filter in regards to domain specific technical terms.}
\extended{One could also train models with a wider range of parameters, some that we touched like the number of aspects and different datasets, as well as some that we did not cover like experimenting with the (word) embedding size or trying a different optimizer.}
\extended{The evaluation could also be expanded to calculate the output distribution of the models to check if all outputs are used to an equal amount.
Another direction of research could also be to exploit current Transformer models~\cite{BERT}, which however again will require labeled training data.}

\extended{Beyond the discussion of main results, threats to validity, generalization, and practical impact and future work, we also conducted an extensive retrospective analysis of the lessons learned during our research.
These lessons learned are reported as Appendix~\ref{sec:lessonslearned}.}

\section{Conclusion}
\label{sec:conclusion}
STEREO is a tool to analyze and extract sentences containing statistics from scientific papers.
We have shown that finding and extracting sentences containing statistics with our hierarchical regex-based wrapper works very well for both APA-conform and non-APA reports.
The extraction precision of experiment conditions and experiment topic between $30\%$ to $45\%$ is reasonable given the variety and challenging especially in non-APA conform reports.
\extended{These results could be improved as described in the future work above.
\section*{Data availability and Reproducibility}}
The source code of STEREO, rule sets, and models are available here: \url{https://github.com/Foisunt/STEREO}.

\extended{
\section*{Acknowledgments}
We thank Johannes Keller for providing us his anglistics knowledge for the grammar-based patterns for extracting experiment conditions.
We thank Jessica Töllich and Lukas Galke for feedback on this manuscript.

The presented research is the result of a Master module ``Project Data Science'' taught at the University of Ulm in summer term 2020.
The last author is supervisor of the student group.}

\bibliographystyle{ACM-Reference-Format}
\bibliography{acmart}

\extended{
\newpage

\appendix

\section{Details of ABEA for Topic Extraction}
\label{sec:detailsofabea}
We provide a brief technical description of ABEA~\cite{abae} as used in our context for topic extraction.
First the vocabulary size $V$ and the word embedding dimension $d$ of the underlying word embedding as well as the number of different aspects $K$ have to be chosen.
The $K$ aspects represent in our case the $K$ different experiment topics that may be covered by an input sentence.

\begin{figure}[h]
    \centering
    \includegraphics[width=0.48\textwidth]{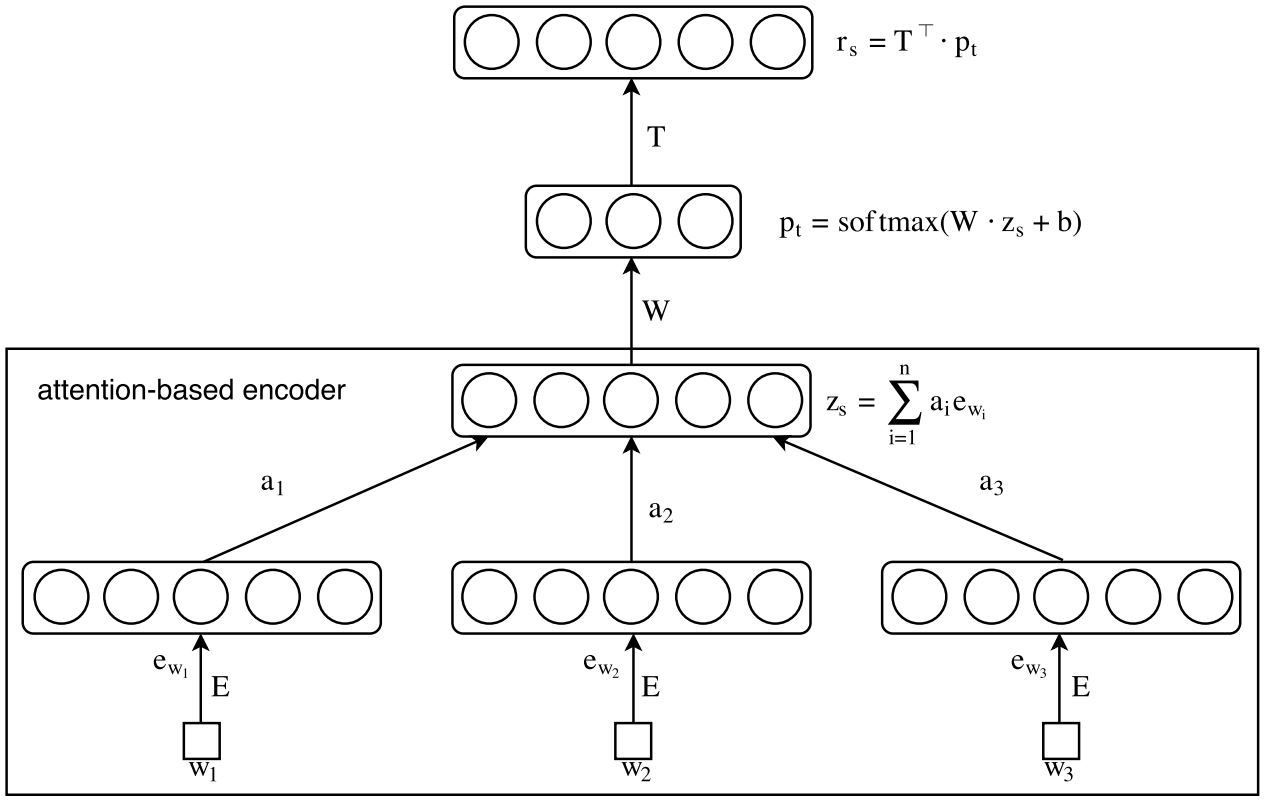}
    \caption{ABAE structure, image taken from \cite{abae}.}
    \label{fig:abae}
\end{figure}

The input to the ABEA model is a sentence or set of words 
$ s =\{  w_1, \ldots, w_n\}$.
Each word corresponds to a row of the embedding matrix $E \in \mathds{R}^{V\times d}$ and can be transformed to a vector $e_{w_i} \in \mathds{R}^{d}$. 
This word embedding describes the local context of the word.
Figure~\ref{fig:abae} shows the whole architecture of ABAE, with this step corresponding to the bottom most arrows in the figure.
Equation~\ref{eq:z_s} shows how the sentence embedding $z_s$ is computed from the word embeddings. 
This can also be seen in the center of \mbox{Figure~\ref{fig:abae}}.

\begin{equation}
z_s = \sum_{i=1}^n a_i \cdot e_{w_i}
\label{eq:z_s}
\end{equation}

The weights $a_i$ describe how important a word is for the meaning of the sentence.
They are calculated by the attention mechanism shown in Equations~\ref{eq:a_i} to~\ref{eq:y_s}.
$y_s \in \mathds{R}^d$ is the average word embedding of a sentence and describes the global context of a sentence.
The matrix $M \in \mathds{R}^{d\times d}$ in Equation~\ref{eq:d_i} is a mapping between the global and local context.

\begin{align}
    \label{eq:a_i}
    a_i &= \frac{exp(d_i)}{\sum_{j=1}^n exp(d_j)}\\
    \label{eq:d_i}
    d_i &= e_{w_i}^T\cdot M\cdot y_s\\
    \label{eq:y_s}
    y_s &= \frac{1}{n}\sum_{i=1}^n e_{w_i}
\end{align}

On top of the sentence embedding is an auto-encoder like structure, which can also be seen in the top part of Figure~\ref{fig:abae}. 
The aspect probability vector $p_t$ gets calculated like shown in Equation~\ref{eq:p_t} with $W \in \mathds{R}^{K\times d}$ and $b \in \mathds{R}^{K}$.

\begin{equation}
    \label{eq:p_t}
    p_t = \text{softmax}(W\cdot z_s + b)
\end{equation}

Finally the sentence embedding is reconstructed from the aspect probability vector with the aspect embedding matrix $T \in \mathds{R}^{Kxd}$. 

\begin{equation}
    r_s = T^t \cdot p_t
\end{equation}

The aim while training ABAE is to minimize the reconstruction error between $r_s$ and $z_s$, and to maximize the difference between $r_s$ and the average word embedding of any negative sample $n_i$.
A negative sample is a sentence from the input data with a different aspect than the current sentence $s$.

As ABAE is unsupervised neither the aspect of the current sentence nor the aspect of any other sentence is known before and during training, the $m$ negative samples are randomly drawn from the input data for each sentence and over multiple training epochs. Most negative samples should have had a different aspect than $s$.
To implement this, a modified hinge loss is used that is proportional to $r_sn_i$ and negative proportional to $r_sz_s$ as shown in Equation~\ref{eq:J}.

To improve diversity of the learned aspects, a regularization $U$ that promotes orthogonality of the aspect embedding matrix's rows is added.
$T_n$ is the matrix $T$ with each row normalized to length $1$ and $I$ is the identity matrix.
This regularization and it's combination with the hinge loss to the overall loss $L$ are shown by the following equations:

\begin{align}
    \label{eq:J}
    J(\theta) &= \sum_{s\in D} \sum_{i=1}^m max(0,1-r_sz_s+r_sn_i)\\
    U(\theta) &= \Vert T_n \cdot T_n^t - I \Vert \\
    L(\theta) &= J(\theta)+\lambda U(\theta).
\end{align}

Finally, the most representative words of each aspect are extracted from the word and aspect embeddings and the aspects are manually inferred from those.
For example in the original paper ``\textit{main dishes}'' was inferred from the representative words ``\textit{beef, duck, pork, mahi, filet, veal}'' and ``\textit{dessert}'' from ``\textit{gelato, banana, caramel, cheesecake, pudding, vanilla}''.
``\textit{main dishes}'' and ``\textit{dessert}'', which where then mapped to the gold standard aspect ``\textit{Food}'' as they had 14 inferred for 6 gold standard aspects.
A sentence gets assigned one of those inferred aspects according to the aspect probability vector from Equation~\ref{eq:p_t}.

\section{Lessons Learned}
\label{sec:ll}
\label{sec:lessonslearned}
One problem that appeared in the statistic extraction learning phase for our rules is, that some sentences contained syntactic deviations from the original papers, which probably happened due to parsing errors from PDF to JSON.
Problematic is that the kinds of parsing errors differ from paper to paper.
Some errors are caused, because a single uni-code character could not be correctly translated.
Some of these errors appeared often and could potentially have an impact on detecting statistics.
One of these errors was that a lower case \texttt{L} (l) has been parsed into the digit 1.
In the statistic notations from our supported statistics, we would not expect a lower case \texttt{L} at all and especially not instead of a numerical value.
Therefore, if this parsing error would happen in a statistical record, the sentence would not be detected as $R^+$.
However, while learning the rules with the wrapper, we did not find any statistics containing such a parse error.

Another error that appeared in statistical records is that for negative values, the minus character (-) would not match, because through the parsing process the character has been transformed into another Unicode character which looks almost the same.
However, we were able to fix this issue by allowing alternative variants of the - in their respective Unicode encoding.
Furthermore, in some documents, the citation syntax could not be properly parsed, so that instead of a digit inside a bracket \textit{[...]}, e.\,g. \textit{[7]}, only the digit would be printed.
The digit was then concatenated to the beginning of the next sentence, so that our pattern for splitting sentences could not detect the ending of the sentence, since we normally do not expect a sentence to start with a digit.
We also observed a similar behavior, when the original papers did contain line numbers.
Overall, this issue could be technically challenged, but it is very cumbersome and labor intensive.

In the wrapper induction approach for learning statistical records, after loading a document, it will be checked in which language the paper is written.
If another language than English is detected, the document will be skipped.
This is done using the python library langdetect.
This routine does not work perfectly, i.\,e. ''Viruses are unique in nature'' will be detected as French with a confidence of$>99.9\%$.
However, in general the language detection tool has a high accuracy $0.845$. \footnote{According to \url{https://towardsdatascience.com/benchmarking-language-detection-for-nlp-8250ea8b67c}}.
Thus, we do not assume that this is a large problem.

For some sentences containing a statistical record,  it is not possible to extract the respective conditions, e.\,g.
''\textit{The subjects were found to be significantly (t (263) = 25.04, p<0.001)}''.
However the original sentence as it is contained in the CORD-19 dataset is
''\textit{The subjects were found to be significantly (t (263) = 25.04, p<0.001) overweight (23.01 ± 16.82 kg) and had a mean excess of 4.95± 16.82 kg of fat.}''.
We made the assumption that the record is located at the end of the sentence, which would be conform with APA style. 
After a statistic is found, we only extract the part of the sentences up to where the statistic is located.
However, in the original sentence, the aspect is located directly after the statistical record and thus not present in the sentence we have extracted.
Although every variables of the statistical record of a $t$-test are found (i.\,e., the $t$-value, df, $p$-value ...), this sentence may not contain the information about the experiment conditions or experiment topic.
With our current model, we do not preserve this contextual information, since we work on a single sentence level the preceding and succeeding sentences are not considered.

Specifically for GBCE, it is sometimes not enough to have one sentence as input, since the context is not clear without the additional information provided by neighboring sentences.
For example, in the sentence ``\textit{Increased number of OUCC patients received antibiotics within $60$ minutes after algorithm implementation}'', it is  not clear without contextual information, if ``OUCC patients'' is a condition or if the full example is an aspect.
Furthermore, a distinction in the results between different statistical test types could be of value to further increase the certainty of a high generalizability of GBCE.
}
\end{document}